\documentclass[11pt]{article}

\usepackage[a4paper,margin=1in]{geometry}
\usepackage[T1]{fontenc}
\usepackage[utf8]{inputenc}
\usepackage{lmodern}
\usepackage{microtype}
\usepackage{graphicx}
\usepackage{amsmath}
\usepackage{enumitem}
\usepackage[numbers,sort&compress]{natbib}
\usepackage{xurl}
\usepackage[colorlinks=true,linkcolor=blue,citecolor=blue,urlcolor=blue]{hyperref}

\setlist[itemize]{leftmargin=*,noitemsep,topsep=3pt}
\newcommand{\keybox}[1]{%
  \begin{center}
  \begin{minipage}{0.92\linewidth}
  \hrule\vspace{0.55em}\small #1\vspace{0.55em}\hrule
  \end{minipage}
  \end{center}}
\newcommand{\leftHighlight}[1]{\keybox{#1}}
\newcommand{\rightHighlight}[1]{\keybox{#1}}

\title{The Era of Extremely Large Optical Telescopes II:\\
The Giant Magellan Telescope and the Thirty Meter Telescope}
\author{Priya Hasan\\
\small Department of Physics, Maulana Azad National Urdu University, Hyderabad, India\\
\small ORCID: 0000-0002-8156-6940}
\date{March 2026}

\begin{document}
\maketitle

\begin{abstract}
The advent of Extremely Large Telescopes (ELTs) - ground-based optical and infrared observatories with primary mirrors larger than 20~m - marks a transformative era in observational astronomy. This article focuses on two of the next generation of optical/infrared facilities: the Giant Magellan Telescope (GMT) and the Thirty Meter Telescope (TMT). It summarizes the technological ideas that make these observatories possible, especially segmented or multi-mirror primary apertures, active optics, adaptive optics, laser guide-star systems, and high-throughput instrumentation. Together, these developments will provide a major increase in light-collecting area and angular resolution over present facilities. In diffraction-limited adaptive-optics modes, the ELTs will reach image sharpness beyond that of current space telescopes at comparable infrared wavelengths, although space observatories will remain essential for wavelengths blocked by Earth's atmosphere and for stable, low-background observations. The article also discusses why ground- and space-based astronomy are complementary, why satellite interference cannot be solved simply by moving astronomy into space, and why large ground-based observatories will continue to play a leading role in astrophysics.
\end{abstract}

\section{A brief history of telescope apertures}

\leftHighlight{\textbf{A brief history of telescopes}
\begin{itemize}
\item The refractor era (1609--c. 1750)
\item The great refractor and early reflector era (c. 1750--1910)
\item The modern glass-mirror reflector era (c. 1900--1990)
\item The segmented-mirror and adaptive-optics era (c. 1990--2010)
\item The Extremely Large Telescope era (2010s--future)
\end{itemize}}

The pursuit of larger apertures - the diameters of the primary light-gathering surfaces of telescopes - has shaped the history of observational astronomy. A larger aperture collects more light, allows fainter sources to be detected, resolves finer detail, and extends observations to earlier epochs in cosmic history \citep{Racine2004}. This article follows that development from early telescopes to the coming era of Extremely Large Telescopes (ELTs).

The major stages in telescope development can be summarized as follows: the refractor era, the great refractor and early reflector era, the modern glass-mirror reflector era, the segmented-mirror and adaptive-optics era, and the emerging ELT era. Today, the largest practical monolithic optical mirrors are about 8--8.4~m in diameter, as used in facilities such as the Large Binocular Telescope and the Vera C. Rubin Observatory. Beyond this scale, segmented or multi-mirror designs are the viable path to much larger apertures.

\rightHighlight{\textbf{Extremely Large Telescopes (ELTs)} are ground-based observatories with primary apertures exceeding 20~m. Three major optical/infrared facilities define the coming generation:
\begin{itemize}
\item \textbf{Giant Magellan Telescope (GMT):} seven 8.4~m mirrors forming a 25.4~m light-collecting surface.
\item \textbf{Thirty Meter Telescope (TMT):} 492 hexagonal mirror segments forming a 30~m primary mirror.
\item \textbf{European Extremely Large Telescope (ELT):} a 39~m segmented telescope under construction by the European Southern Observatory.
\end{itemize}}

This realization has led to the ELT era - ground-based observatories with primary apertures well above 20~m \citep{Hook2019}. Three facilities are central to this transition: the GMT at Las Campanas Observatory in Chile, the TMT with its preferred site at Maunakea in Hawaii, and ESO's ELT at Cerro Armazones in Chile. This article discusses the GMT and TMT; a companion article discusses ESO's ELT.

Atmospheric seeing, the image blurring caused by turbulence in Earth's atmosphere, has historically limited ground-based telescopes. The diffraction-limited angular resolution of a telescope is approximately
\begin{equation}
\theta \simeq 1.22\frac{\lambda}{D},
\end{equation}
where $\lambda$ is the observing wavelength and $D$ is the aperture diameter. Space telescopes such as the Hubble Space Telescope (HST) and the James Webb Space Telescope (JWST) naturally avoid atmospheric seeing. Space also provides access to ultraviolet, X-ray, and far-infrared wavelengths that are blocked partly or entirely by the atmosphere. Ground-based observatories, however, can now approach their diffraction limits through active optics and adaptive optics. Adaptive optics uses wavefront sensors and deformable mirrors to correct atmospheric distortion in real time, often producing images comparable to, and in some observing modes sharper than, those from current space telescopes.

\rightHighlight{\textbf{Major revolutions in telescope design}
\begin{itemize}
\item Large monolithic reflectors up to about 8.4~m
\item Segmented or multi-mirror primaries for larger apertures
\item Active optics and adaptive optics
\item Interferometry at radio, millimeter, and optical wavelengths
\item Space telescopes for atmosphere-free observing
\end{itemize}}

There has also been a revolution in angular resolution through interferometry, especially in radio, millimeter, and optical astronomy. These innovations have enabled many of the major discoveries of modern astrophysics. In the following sections, we focus on two next-generation facilities: the Giant Magellan Telescope and the Thirty Meter Telescope.

\section{The Giant Magellan Telescope}

\leftHighlight{The GMT has a distinctive optical design: it combines seven monolithic 8.4~m mirrors into a 25.4~m light-collecting surface rather than using hundreds of small mirror segments.}

The GMT is a next-generation optical/infrared telescope under construction at Las Campanas Observatory in Chile's Atacama Desert \citep{Johns2012,Bernstein2014,GMTODesign}. The official GMT design uses seven of the world's largest mirrors, each 8.4~m in diameter, arranged in a flower-like pattern to form a 25.4~m light-collecting surface \citep{GMTODesign}. This design gives the GMT a large collecting area while retaining many of the advantages of high-quality monolithic optics.

Past giant telescopes, such as the Keck telescopes, used many smaller segments. The GMT takes a different route by using seven large monolithic mirrors. These mirrors are produced at the University of Arizona's Richard F. Caris Mirror Lab. Low-expansion glass is melted in a rotating furnace, forming a lightweight honeycomb mirror blank. After slow cooling, each blank is polished to extremely high precision and coated for astronomical use. The off-axis outer segments are particularly challenging because their surfaces are asymmetric.

\begin{figure}[t]
\centering
\includegraphics[width=0.88\linewidth]{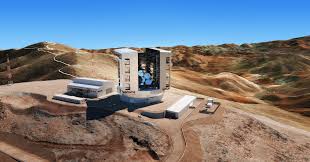}
\caption{Artist's rendering of the GMT enclosure and site at Las Campanas Observatory in Chile. Image credit as supplied in the original manuscript: M3 Engineering/GMTO Corporation.}
\label{fig:gmt_site}
\end{figure}

The GMT's seven mirrors will operate as a single optical system. The primary mirror segments collect light and send it to seven adaptive secondary mirrors. These secondary mirrors correct atmospheric distortions by changing shape rapidly, allowing the telescope to approach its diffraction limit in adaptive-optics observing modes \citep{GMTODesign}. The GMT official design describes adaptive secondary mirrors deforming at roughly 2000 times per second to correct atmospheric turbulence \citep{GMTODesign}.

\begin{figure}[t]
\centering
\includegraphics[width=0.75\linewidth]{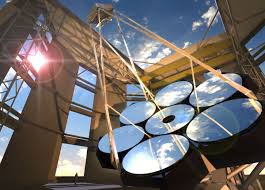}
\caption{Conceptual view of the GMT's seven-mirror primary aperture. The telescope uses seven 8.4~m mirrors to form a 25.4~m light-collecting surface. Image credit as supplied in the original manuscript: GMTO Corporation.}
\label{fig:gmt_mirrors}
\end{figure}

This combination of aperture and adaptive optics will enable a wide range of science. The GMT will study exoplanets and their atmospheres, resolve crowded stellar fields, trace the formation of stars and galaxies, and study black holes and galactic nuclei. Current telescopes can directly image mainly giant planets in wide orbits; ELT-class facilities will push toward fainter, smaller, and closer-in worlds and will allow high-dispersion spectroscopic studies of planetary atmospheres.

\section{The Thirty Meter Telescope}

\leftHighlight{\textbf{Why does TMT use a three-mirror anastigmat?} A classical two-mirror telescope suffers from optical aberrations such as coma, astigmatism, and field curvature. TMT's three-mirror anastigmat design is intended to control these aberrations over a wide field, giving uniform image quality and efficient delivery of light to its science instruments.}

The TMT is designed as a 30~m optical/infrared telescope with a segmented primary mirror made of 492 individual 1.4~m hexagonal segments \citep{Nelson2008,Skidmore2015,TMTOptics,TMTMirrorTracker}. Its international partnership includes institutions in the United States, Canada, Japan, India, and China. The TMT is especially important because it is planned as the major northern-hemisphere ELT, complementing the large southern facilities in Chile. TMT's official public materials continue to identify Maunakea as its preferred site \citep{TMTInstruments}.

\begin{figure}[t]
\centering
\includegraphics[width=0.88\linewidth]{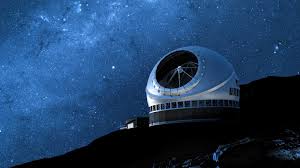}
\caption{Artist's rendering of the Thirty Meter Telescope enclosure. The preferred TMT site is Maunakea, Hawaii, while La Palma in the Canary Islands has been discussed as an alternate site. Image credit as supplied in the original manuscript: TMT International Observatory.}
\label{fig:tmt_enclosure}
\end{figure}

TMT's angular resolution will be much finer than that of JWST at comparable wavelengths because angular resolution scales approximately as $\lambda/D$. JWST has a 6.5~m primary mirror, while TMT's primary mirror is 30~m. In adaptive-optics modes, TMT will therefore be able to separate compact structures that appear blended to smaller telescopes.

TMT's first-light adaptive-optics system is NFIRAOS, the Narrow Field InfraRed Adaptive Optics System. NFIRAOS is a facility-class multi-conjugate adaptive-optics system using two deformable mirrors and six laser guide stars. TMT's official instrument page states that NFIRAOS is designed to provide uniform high-Strehl correction across a one-arcminute field and to feed instruments including IRIS and MODHIS \citep{TMTInstruments}. The first-light and early instruments include:
\begin{itemize}
\item \textbf{WFOS}: the Wide-Field Optical Spectrometer, for imaging and spectroscopy across a large optical field.
\item \textbf{IRIS}: the Infrared Imaging Spectrograph, an imager and integral-field spectrograph fed by NFIRAOS.
\item \textbf{MODHIS}: the Multi-Objective Diffraction-limited High-Resolution Infrared Spectrograph.
\end{itemize}

\begin{figure}[t]
\centering
\includegraphics[width=0.95\linewidth]{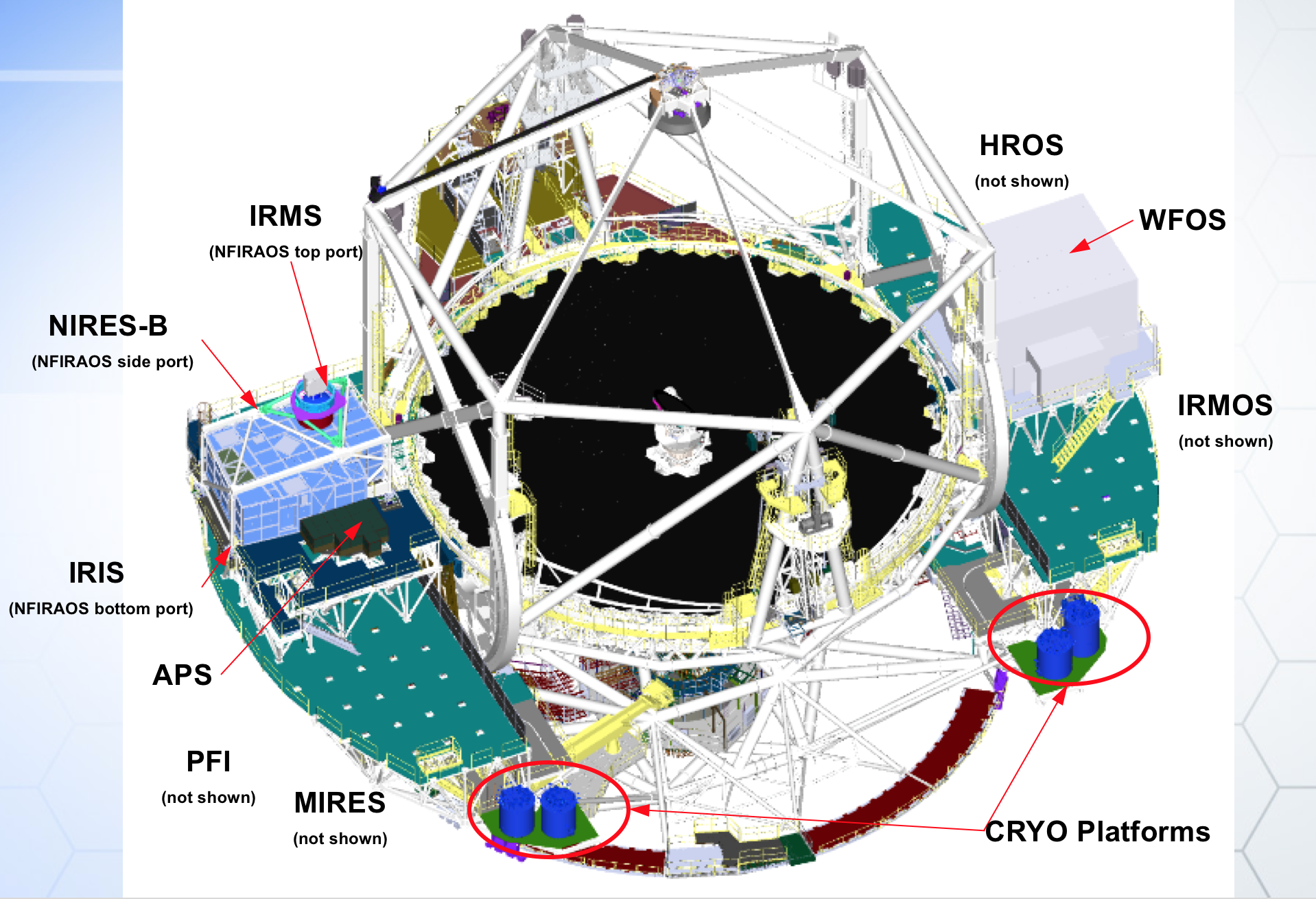}
\caption{Stylized TMT instrument layout. WFOS is shown on the right, while NFIRAOS on the left feeds instruments such as IRIS and MODHIS. Image credit as supplied in the original manuscript: TMT International Observatory.}
\label{fig:tmt_instruments}
\end{figure}

In the near-ultraviolet and blue optical range, TMT aims to offer an important niche through high-throughput optical spectroscopy. WFOS is designed for wide-field optical spectroscopy and for multi-object studies of faint sources. In the near-infrared, the combination of the 30~m aperture and NFIRAOS will make TMT powerful for crowded-field stellar astronomy, black-hole dynamics, high-redshift galaxies, and exoplanet studies. In the mid-infrared, the cold, dry conditions of high-altitude sites are important because the thermal background from the atmosphere and telescope becomes a limiting factor.

\rightHighlight{\textbf{TMT construction and instrumentation update}
\begin{itemize}
\item TMT's 30~m primary mirror is designed to use 492 hexagonal mirror segments, with additional spare segments produced for operations.
\item TMT public materials report active production of mirror blanks, roundels, and polished mirror assemblies across partner institutions.
\item NFIRAOS, IRIS, MODHIS, and WFOS define much of the early science capability.
\item The project timeline remains closely tied to final site, funding, and construction decisions.
\end{itemize}}

The ELT, GMT, and TMT are therefore not simply scaled-up versions of existing observatories. Each has a different balance of aperture, site, optical design, wavelength coverage, instrument suite, and adaptive-optics strategy. Together, they will form a complementary global system.

\section{The TMT site controversy}

The TMT site controversy presents the ethical and social dimensions of building very large observatories. A telescope of this scale cannot be built just anywhere. It requires exceptionally clear, dark, dry, and stable skies, as well as infrastructure capable of supporting a complex international facility.

\leftHighlight{\textbf{Why Maunakea has been important for astronomy}
\begin{itemize}
\item High altitude, dry air, and excellent atmospheric stability
\item Dark skies and good infrared observing conditions
\item Existing astronomical infrastructure
\item Major facilities already present, including Keck, Subaru, Gemini North, CFHT, IRTF, UKIRT, JCMT, and the Submillimeter Array
\end{itemize}}

For astronomers, Maunakea offers exceptional observing conditions. For many Native Hawaiians, however, Maunakea is sacred: a place of cultural, spiritual, ancestral, and ecological significance. The conflict is therefore not merely a technical question of site selection. It is also about history, governance, land use, cultural respect, and indigenous sovereignty.

Since the 1960s, many telescopes have been constructed on Maunakea. Critics argue that earlier development often proceeded without adequate cultural and environmental accountability, creating a deep lack of trust. The TMT, because of its size and visibility, became a focal point for these broader concerns. Proponents emphasize the scientific value, educational and economic benefits, and attempts to develop more responsible management practices. Opponents, including many Native Hawaiian protectors (kiai), view the project as another intrusion on a sacred mountain.

Legal challenges and protests, especially in 2015 and 2019, have shaped the project's timeline. Hawaii's Supreme Court upheld the construction permit in 2018, but the social and political conflict remains unresolved. The alternate site most often discussed is the Roque de los Muchachos Observatory on La Palma in Spain's Canary Islands, which is already home to major astronomical facilities. Although La Palma would not reproduce all of Maunakea's advantages, it could still support much of TMT's core science.

\rightHighlight{\textbf{TMT site options discussed in project history}
\begin{itemize}
\item Maunakea, Hawaii, United States
\item Roque de los Muchachos Observatory, La Palma, Canary Islands, Spain
\item Cerro Armazones, Chile
\item Cerro Tolanchar, Chile
\item Cerro Tolar, Chile
\item San Pedro Martir, Baja California, Mexico
\item Hanle, Ladakh, India
\end{itemize}}

The debate is not a simple binary of science versus culture. For Maunakea, the scientific case includes exceptional observing conditions and northern-sky coverage. For La Palma, the case includes a scientifically capable site with a different social and legal context. The central issue is whether the long history of mistrust on Maunakea can be addressed through a genuinely respectful partnership, or whether relocation is the more responsible path. The TMT story reminds us that science operates within society, and that the pursuit of knowledge must be balanced with cultural, environmental, and ethical responsibilities.

\section{Conclusion: the enduring value of ground-based astronomy}

Space-based astronomy is not a new dream. Astronomers have considered orbital observatories for generations, and space telescopes have transformed astrophysics. Yet the astronomical community has not moved entirely into space, and there are compelling reasons why it cannot.

Space telescopes are specialized instruments. They avoid atmospheric distortion and atmospheric absorption, and they can observe wavelength ranges inaccessible from the ground. However, they are expensive, limited in aperture by launch constraints, difficult or impossible to repair, and usually fixed in capability after launch.

Ground-based observatories are different. They are long-lived institutions. Their instruments can be replaced, mirrors can be recoated, detectors can be upgraded, and new capabilities can be added over decades. The ELTs - the GMT, TMT, and ESO's ELT - are possible precisely because they are built on Earth. A 25--40~m class optical/infrared telescope cannot be launched into space with present technology.

\leftHighlight{\textbf{Ground-based telescopes}
\begin{itemize}
\item Pros: larger apertures, easier maintenance, lower cost per unit aperture, and upgradeable instruments.
\item Cons: atmospheric distortion, weather losses, atmospheric absorption, light pollution, radio interference, thermal effects, and gravity-dependent distortions.
\end{itemize}
\textbf{Space-based telescopes}
\begin{itemize}
\item Pros: no atmospheric seeing, access to otherwise blocked wavelengths, stable environment, and low sky background in some bands.
\item Cons: size limits, high cost, limited repair options, finite mission lifetimes, and launch risk.
\end{itemize}}

The rise of satellite constellations has created a serious challenge for ground-based astronomy. However, moving all astronomy into space is not a solution. Satellite trails, radio-frequency interference, orbital congestion, and space debris are policy and engineering problems. They require regulation, coordination, and responsible design, not the abandonment of ground-based observing.

The coming ELT era is therefore not a replacement of one observing platform by another. It is a networked future. The GMT, TMT, ESO's ELT, the Vera C. Rubin Observatory, JWST, the Nancy Grace Roman Space Telescope, and many other facilities will work together. Space telescopes will provide stable, atmosphere-free views; ground-based ELTs will provide enormous apertures, adaptive-optics resolution, high-resolution spectroscopy, and decades of upgradeable observing power. Their combined capabilities will open discoveries that no single facility could achieve alone.

\section*{Acknowledgements}
The author thanks colleagues and reviewers whose comments helped improve the clarity and balance of this article.

\end{document}